\documentclass[nofootinbib,aps,preprint,onecolumn,amsmath,amssymb,superscriptaddress,eqsecnum,floatfix,scrartcl]{revtex4-1}
\pdfoutput=1      
\usepackage{amssymb}  
\usepackage{color,soul} 
\usepackage{graphicx}  
\usepackage[tight]{subfigure}  
\usepackage{mathrsfs}
\usepackage{bbm}
\usepackage[pdftex,pdfusetitle,bookmarks=true,colorlinks=true,citecolor=blue,urlcolor=blue,linkcolor=magenta]{hyperref}
\usepackage{hypcap}

\begin{document}

\title{On conformal perturbation theory}    
\author{Andrea Amoretti}   
\email{andrea.amoretti@ulb.ac.be} 
\affiliation{Physique Th\'{e}orique et Math\'{e}matique Universit\'{e} Libre de Bruxelles, C.P. 231, 1050 Brussels, Belgium}

\author{Nicodemo Magnoli} 
\email{nicodemo.magnoli@ge.infn.it} 
\affiliation{Dipartimento di Fisica, Universita di Genova, Via Dodecaneso 33, 16146, Genova, Italy}
\affiliation{INFN, sezione di Genova, Via Dodecaneso 33, 16146, Genova, Italy}

\date{\today}

\begin{abstract}
Statistical systems near a classical critical point have been intensively studied both from theoretical and experimental points of view.
In particular, correlation functions are of relevance in comparing theoretical models with the experimental data of real systems. In order to compute physical quantities near a critical point one needs to know the model at the critical (conformal) point. In this line, recent progresses in the knowledge of conformal field theories, through the conformal bootstrap, give the hope to get some interesting results also outside of the critical point. In this note we will review and clarify how, starting from the knowledge of the critical correlators, one can calculate in a safe way their behavior outside the critical
point. The approach illustrated requires the model to be just scale invariant at the critical point. We will clarify the method by applying it to different kind of perturbations of the $2D$ Ising model.
\end{abstract}

\maketitle   
\tableofcontents      

\section{Introduction}
\label{Intro}
Recently there has been a renewed interest in conformal field theories (CFT) in higher dimensions. Some new insight has been obtained through the conformal bootstrap method \cite{ElShowk:2012ht,El-Showk:2014dwa, Gliozzi:2013ysa,Gliozzi:2014jsa,Kos:2015mba}. One of the most important results has been the computation of the critical exponents and the Wilson coefficients of the $3D$ Ising model \cite{El-Showk:2014dwa, Gliozzi:2013ysa}. The knowledge of the system at the critical point gives us the possibility to analyze its behavior away from criticality, considering \emph{e.g.} the perturbations due to the presence of an external magnetic field or an energy operator. The knowledge of the correlators in the vicinity of the critical point is extremely relevant to compare theoretical predictions with experimental data, to study, in the $2D$ case, the relation between perturbed CFTs and integrable field theories \cite{Zamolodchikov:1989zs} or the flow between different CFTs \cite{Zamolodchikov:1987ti}.
From the theoretical point of view, to obtain the behavior of observables outside a critical point is not an easy task, since usually the standard approach based on the Gell-Mann and Low theorem \cite{PhysRev.84.350} is not useful in this situation. This is due to the fact that, in the case of a conformal field theory   perturbed by a relevant operator, the corrections to the conformal correlation functions are typically plagued by infrared divergences. This problem has been firstly pointed out by Wilson in its seminal paper \cite{Wilson:1969zs}. 
In order to understand the issue and its possible solution it is useful to consider the theory of the free massive boson. It is well known that the euclidean propagator of this theory can be expressed in terms of a modified Bessel function depending on the product of the mass $m$ and the distance $r$ (times a power of $r$). The short distance 
($m r << 1$)  behavior of the Bessel function contains terms proportional to $Log (m r)$, which cannot be obtained by performing a small mass perturbation at any finite order. Additionally, naive perturbative mass expansion is affected by infrared divergences, being the massless ``free'' theory scale invariant. It is clear that the correlation function by itself has no  finite perturbative expansion. As pointed out in \cite{Wilson:1969zs}, one can overcome the problem by  expressing the short distance behavior of the correlators in terms of the Operator Product Expansion (OPE). In this way, the non perturbative contribution of the correlators, which is encoded in the vacuum expectation values (VEV) of the local operators in the OPE, separates from the one that can be computed perturbatively, namely the Wilson coefficients. Eventually, if the expectation values of the operator are known, the problem reduces to perturbatively compute the Wilson coefficients.  

Even though techniques to perturbatively compute the short distance IR finite behavior of the Wilson coefficients have been developed a long time ago \cite{Guida:1995kc}, it seems that the method is not well known in the recent literature. In this paper we will review the approach firstly described in \cite{Guida:1995kc}, which can be applied  to compute the behavior of the correlators of models of arbitrary space dimensions in the vicinity of  a scale invariant critical point. In order to render the method as clear as possible, we will apply it to several example of perturbations of the 2D Ising model which have not yet been considered in the literature\footnote{See \emph{e.g.} \cite{Caselle:1999mg,Caselle:2001zd,Caselle:2003ad} for other examples of perturbations of the 2D Ising model and \cite{Zamolodchikov:1990bk} for an application to the  Lee-Yang model.}. Even though the purpose is pedagogical, each example has a physical interest by itself, which will be discussed in details in the conclusions.

The paper is organized as follows. In Section 2  we describe the general method to perturbatively compute the Wilson coefficients, clarifying some aspects not discussed in \cite{Guida:1995kc}. In section 3 we apply the method to three examples related to the $2D$ Ising model.  Firstly, we consider the correlators of the Ising model perturbed by the energy operator $\mathcal{E}$ in the broken phase, where the spin operator $\sigma$ has a non-trivial expectation value. Specifically, we will compare the perturbative result for the correlator $\langle \sigma \mathcal{E} \rangle$ with the analytical expression obtained by Hecht \cite{PhysRev.158.557}. This provide us with an example in which conformal perturbation theory gives the correct results even in a phase where the underlying symmetry of the model is somehow broken. Secondly we apply the method to the Ising model perturbed by the magnetic field in the presence of a trap. This example might be of relevance to compare theoretical predictions with data from experimental setups in which the presence of a trap is needed to confine the system in a limited region, such as Bose Einstein condensates and cold atoms \cite{trap1,trap2,trap3,trap4}. Finally we will show that the method can be used to consistently compute perturbations to the three point functions as well. 
In this case we will show that the three-point correlation function satisfies an associativity condition outside the critical point that can give non trivial relations 
between the Wilson coefficients and its derivatives.
The last Section of the paper is devoted to conclusions and final remarks.

\section{Conformal perturbation theory: a review}
\label{CPT}
Let us review the technique derived in \cite{Guida:1995kc}, to reconstruct the short distance behavior of the correlators of operators in a $D$ dimensional Euclidean quantum field theory at a fixed point of the renormalization group, described by the action $S_e$, perturbed by one ore more relevant operators $\mathcal{O}_i$ with couplings $m_i$, so that the perturbation to $S_e$ takes the form:
\begin{equation}\label{perturb}
\Delta S =- \int d^D x \ \sum_i m^i \ \mathcal{O}_i(x) \ .
\end{equation}
Since the operators $\mathcal{O}_i$ are supposed to be relevant, their dimensions $x_i$ have to satisfy the inequality:
\begin{equation}
0<x_i<D \ .
\end{equation}
In what follows, we will be especially interested in statistical physics applications. In this spirit, the unperturbed theory described by $S_e$ is a conformal field theory, and the method which we are going to outline describe a consistent way to go outside the critical point.

In general, the short distance behavior of the correlators of a perturbed field theory is described by the operator product expansion (OPE):
\begin{equation}\label{OPE}
\langle \Phi_{a_1}(r_1) ... \Phi_{a_n}(r_n) \ X \rangle_m \sim C^c_{a_1...a_n}(r_1-r_n,... \ ,r_{n-1}-r_n \ ; \ m) \ \langle\Phi_c(r_n) \ X(R) \rangle_m \ ,
\end{equation}
where $\Phi_i(r_i)$ describes a complete set of composite operators of dimension $x_i$, $X(R)$ could be either the identity operator   ($X(R)=\mathbbm{1}$) or a multi-local operator defined on $|R|>|r|, \ ... \ , |r_n|$, and the index $m$ indicates that the correlators are computed at fixed sources $m_i$.

If the unperturbed theory ($m_i=0$) is a conformal field theory, the convergence of the OPE at $m_i=0$ is well understood \cite{PhysRevD.86.105043}. However, the convergence of OPE in a general perturbed field theory is far from being proved. By the way, for the sake of the method to be valid, it is sufficient to assume the asymptotic weak convergence of the OPE, namely that the truncated OPE expansion 
\begin{equation}\label{defdeltan}
\Delta^{(N)}_{a_1 ...a_n}(X(R), \ m ) \equiv \langle \left( \Phi_{a_1}(r_1)... \Phi_{a_n}(r_n)-\sum_c^{x_c \le N} \ C^c_{a_1...a_n}(\{r_i\};\ m) \ \Phi_c(r_n) \right) \ X(R) \rangle_m
\end{equation}
satisfies the condition
\begin{equation}\label{condi}
\lim_{N \rightarrow \infty} \ \lim_{R \rightarrow \infty} \Delta^{(N)}_{a_1 ...a_n}(X(R), \ m ) \sim 0
\end{equation}
as an asymptotic series in the couplings $m_i$. The dependence on the coordinates $\{r_i\}$ of the Wilson coefficients $C^c_{a_1...a_n}$ has to be understood as in \eqref{OPE}. The condition \eqref{condi} is equivalent to the assumption that arbitrary order derivatives of $\Delta^{(N)}_{a_1 ...a_n}$ with respect to the couplings $m_i$ asymptotically converge in the $m_i \rightarrow 0$ limit, namely:
\begin{equation}\label{weakhypo}
\lim_{N\rightarrow\infty} \ \lim_{m \rightarrow 0} \ \lim_{R \rightarrow \infty} \partial_{m_1}^{n_1} ... \partial_{m_k}^{n_k} \Delta^{(N)}_{a_1 ...a_n}(X(R), \ m )=0
\end{equation}
for every $k$ and $n_i$. 

\subsection{First order expansion}
Having described the basic hypothesis on which we rely, we are now ready to describe the method. For the sake of simplicity, let us firstly concentrate on the first order expansion. The goal will be to express the first derivative of the Wilson coefficients $C^c_{a_1...a_n}$ with respect to a certain coupling $m_i$ in terms of quantities of the unperturbed theory. 

From the weak convergence hypothesis \eqref{weakhypo}, we know that:
\begin{equation}\label{firstorde}
\lim_{N\rightarrow \infty} \ \lim_{m \rightarrow 0} \ \lim_{R \rightarrow \infty}\partial_{m_i} \Delta^{(N)}_{a_1...a_n}(X(R), \ m)=0 \ .
\end{equation}
In order to expand the previous expression in a useful way, we notice that, due to the basic action principle \cite{Schwinger:1951xk,Schwinger:1953tb}, the following equality holds:
\begin{equation}\label{actionprinciple}
\partial_{m^i}  \langle X(R) \rangle_m=\int d^Dx \ \langle : \mathcal{O}_i(x) : \ X(R) \rangle \  , \text{ with } : \mathcal{O}_i(x) : \ \equiv \mathcal{O}_i- \langle \mathcal{O}_i \rangle \ .
\end{equation}
Keeping in mind the definition of $\Delta^{(N)}_{a_1...a_n}(X(R), \ m)$ \eqref{defdeltan}, and using \eqref{actionprinciple}, equation \eqref{firstorde} becomes:
\begin{multline}\label{con1}
\lim_{N\rightarrow \infty} \lim_{R \rightarrow \infty} \left[ \int_{|\bar r|<|R|}d^D\bar r \ \langle : \mathcal{O}_i(\bar r) : \left(\Phi_{a_1}(r_1)...\Phi_{a_n}(r_n)-\sum_c^{x_c \le N} C_{a_1...a_n}^c(\{r_i\}) \Phi_c(r_n) \right) X(R) \rangle \right.\\
\left.-\sum_{c}^{x_c \le N} \partial_{m_i} C^c_{a_1...a_n}(\{r_i\}) \langle \Phi_c(r_n) X(R)\rangle\right]=0 \ ,
\end{multline}
where the Wilson coefficients $C^c_{a_1...a_n}(\{r_i\})$ and the expectation values $\langle...\rangle$ without the index $m$ are the ones evaluated at $m_i=0$. The relation \eqref{con1} establishes a set of constraints which involve the first derivative of the Wilson coefficient $C_{a_1...a_n}^c$ with respect to $m_i$ and correlators of the unperturbed field theory. Eventually, if one knows the correlators and the Wilson coefficients of the model at the critical point, the system \eqref{con1} can be solved in terms of $\partial_{m_i} C^c_{a_1...a_n}$ obtaining the first order $m_i$ expansion of the Wilson coefficients.

To clarify further equation \eqref{con1}, it is worth to note that it can be further simplified if one considers $X(R)= \mathbbm{1}$, and noticing that, due to dimensional consideration,
\begin{equation}
\lim_{|R_1|...|R_k| \rightarrow \infty} \int_{|r_1|<|R_1|} d^D r_1...\int_{|r_k|<|R_k|} d^Dr_k \langle : \mathcal{O}_i(r_1):... : \mathcal{O}_k(r_k): \Phi_c(r_n) \rangle=0
\end{equation}
if the dimension $x_c$ of $\Phi_c$ is less the the sum of the mass dimensions $y_i \equiv D-x_i$ of each operator $\mathcal{O}_i$, namely if
\begin{equation}
x_c-\sum_{j=1}^k y_j>0 \ .
\end{equation}
This is due to the absence of a physical scale in the unperturbed theory $S_e$. With this in mind, the sum in the second term in \eqref{con1} reduces to all the $\Phi_c$ with $x_c<y_i$, while the only non trivial contribution to the sum in the third term is the one with $\Phi_c(r_n)=\mathbbm{1}$. Finally, one obtains:
\begin{multline}
\partial_{m_i} C^{\mathbbm{1}}_{a_1...a_n}(\{r_i\})=\\
\lim_{R \rightarrow \infty}  \int_{|\bar{r}|<|R|}d^D \bar r \ \langle : \mathcal{O}_i(\bar r) : \left(\Phi_{a_1}(r_1)... \Phi_{a_n}(r_n)-\sum_c^{x_c \le y_i} C_{a_1...a_n}^c(\{r_i\}) \Phi_c(r_n) \right) \rangle \ .
\end{multline}
At this point one might wonder if the previous results might be plagued by ultraviolet divergences. It is worth to stress that in the CPT of a generic CFT there are no issues concerning ultraviolet divergences, as pointed out in \cite{Guida:1995kc}. On the contrary, the free boson case considered in the introduction does not belong to this category and one has to be careful in defining the perturbing composite  operator $\phi ^2$ (see \cite{Guida:1995kc} for a similar discussion in the 2D Ising model and  the example
IIIA in this article). In this situation there appear logarithmic terms  in the short distance expansion.
\subsection{Higher order expansion}
In a similar way, one can consider the expansion of the Wilson coefficients up to higher orders in the derivatives with respect to the couplings $m_i$. By expanding the relation
\begin{equation}
 \lim_{N\rightarrow \infty} \ \lim_{R \rightarrow \infty} \ \lim_{m \rightarrow 0} \partial_{m_1}... \partial_{m_k} \Delta_{ab}^{(N)}(X_R, \ m) \ ,
\end{equation}
and repeating the same steps done in the previous Section, one obtains:
\begin{align} \label{con2}
\lim_{|R|\rightarrow \infty} &\Big\{ \int_{|r_1|<|R|}d^Dr_1 ... \int_{|r_k|<|R|}d^Dr_k   \Big[\langle :\mathcal{O}_{i_k}:...:\mathcal{O}_{i_1}:\Big(\Phi_{a_1}(r_1)...\Phi_{a_n}(r_n)  \nonumber\\
&\qquad \qquad \qquad \qquad \qquad \qquad \; \; \;-\sum_b^{x_b\le\bar{x}}C^b_{a_1...a_n}(\{r_i\})\Phi_b(r_n)\Big)X(R)\rangle\Big] \nonumber\\
&-\sum_{b}^{x_b\le \bar{x}}\partial_{i_1} C^b_{a_1...a_n}(\{r_i\})\int_{|r_2|<|R|}d^dr_2...\int_{|r_k|<|R|}d^Dr_k\langle :\mathcal{O}_{i_k}:...:\mathcal{O}_{i_2}:\Phi_b(r_n)X(R) \rangle \nonumber\\
&... \nonumber \\
&-\sum_b^{x_b\le \bar{x}}\partial_{i_1}...\partial_{i_{k-1}}C^b_{a_1...a_n}(\{r_i\})\int_{|r_k|<|R|}\langle:\mathcal{O}_{i_k}: \Phi_b(r_n)X(R)\rangle \nonumber\\
&-\sum_{b}^{x_b \le \bar{x}} \partial_{i_1} ...\partial_{i_k} C^b_{a_1...a_n}(\{r_i\}) \langle\Phi_b(r_n)X(R) \rangle \Big\} \ ,
\end{align}
where $\bar{x}=\sum_k y_{i_k}-x_{X(R)}$. For generic $X(R)$ the previous relations provide a set of constraints which relate the derivatives of the Wilson coefficients with respect to the coupling to correlators of the unperturbed theory. This system can be consistently solved in terms of the derivatives of the Wilson coefficients if one knows the properties of the model at the critical point.

Once again, we can simplify the previous relation considering $X(R)=\mathbbm{1}$, finally obtaining an equality for the derivatives of $C^{\mathbbm{1}}_{a_1...a_n}$, namely:
\begin{align}\label{confina}
\partial_{i_1}...\partial_{i_k}C^{\mathbbm{1}}_{a_1...a_n}(\{r_i\})=& \nonumber \\
\lim_{|R|\rightarrow \infty} \Big\{ &\int_{|r_1|<|R|}d^Dr_1 ... \int_{|r_k|<|R|}d^Dr_k  \Big[\langle :\mathcal{O}_{i_k}:...:\mathcal{O}_{i_1}:\Big(\Phi_{a_1}(r_1)...\Phi_{a_n}(r_n)  \nonumber\\
&\qquad \qquad \qquad \qquad \qquad \qquad \; \; \;-\sum_b^{x_b\le\bar{x}}C^b_{a_1...a_n}(\{r_i\})\Phi_b(r_n)\Big)\rangle\Big] \nonumber\\
-\sum_{b}^{x_b\le \bar{x}}&\partial_{i_1} C^b_{a_1...a_n}(\{r_i\})\int_{|r_2|<|R|}d^dr_2...\int_{|r_k|<|R|}d^Dr_k\langle :\mathcal{O}_{i_k}:...:\mathcal{O}_{i_2}:\Phi_b(r_n) \rangle \nonumber\\
&... \nonumber \\
-\sum_b^{x_b\le \bar{x}}&\partial_{i_1}...\partial_{i_{k-1}}C^b_{a_1...a_n}(\{r_i\})\int_{|r_k|<|R|}\langle:\mathcal{O}_{i_k}: \Phi_b(r_n)\rangle \Big\} \ .
\end{align}

As a final comment, we note that having obtained the relations \eqref{con1} and \eqref{con2} one has to prove that the expansion is IR finite, namely that all the non perturbative contributions of the correlators are encoded in the expectations values of the operators in the OPE \eqref{OPE}. The proof has been outlined in \cite{Guida:1995kc} and, since it goes beyond the purposes of the present paper, we refer to \cite{Guida:1995kc} for further details.

\section{Examples: 2D Ising model applications}
In this Section we will describe some applications of the conformal perturbation theory techniques previously described to the 2D Ising model. As known (see \emph{e.g.} \cite{Ginsparg:1988ui}), at the critical point, $T=T_c$, the Ising model is described by the continuous unitary conformal field theory $\mathcal{M}(3/4)$. The primary operators of this conformal field theory are $\mathbbm{1}$, $\sigma$ and $\mathcal{E}$, with dimension $x=0,1/8$ and 1 respectively. The corresponding fusion rules are:
\begin{equation}
[\sigma][\sigma]=[\mathbbm{1}]+[\mathcal{E}] \ , \qquad \, [\mathcal{E}][\mathcal{E}]= \mathbbm{1} \ ,\qquad[\sigma][\mathcal{E}]=[\sigma] \ .
\end{equation}
The previous relations implies that correlation functions involving an odd number of $\sigma$s identically vanishes. 

In order to set the conventions and the normalization, we list below some of the correlation functions and Wilson coefficients of this theory which will be useful for our purposes in the following sections:
\begin{equation}\label{wilsoneqsing}
C^{\mathbbm{1}}_{\mathcal{E}\mathcal{E}}(z)=\frac{1}{|z|^2} \ , \qquad C^{\mathbbm{1}}_{\sigma \sigma}(z)=\frac{1}{|z|^\frac{1}{4}} \ , \qquad C^{\mathcal{E}}_{\sigma\sigma}(z)=\frac{|z|^{\frac{3}{4}}}{2} \ , \qquad C^{\sigma}_{\sigma\mathcal{E}}(z)= \frac{1}{2 |z|} \ ,
\end{equation}
and
\begin{equation}\label{3pising}
\langle \sigma(z_1) \sigma(z_2) \mathcal{E}(z_3) \rangle=\frac{|z_{12}|^{\frac{3}{4}}}{2|z_{13}||z_{23}|} \ ,
\end{equation}
\begin{equation}\label{4pising}
\langle \sigma(z_1) \sigma(z_2) \mathcal{E}(z_3) \mathcal{E}(z_4) \rangle=\frac{|z_{12}(z_{32}+z_{42})-2z_{32}z_{42}|^2}{4|z_{42}z_{32}z_{41}z_{31}||z_{43}|^2|z_{12}|^{\frac{1}{4}}} \ ,
\end{equation}
\begin{equation}\label{4pising2}
\langle \sigma(z_1) \sigma(z_2) \sigma(z_3) \sigma(z_4) \rangle=|(1-x) z_{12}z_{34}|^{-\frac{1}{4}} \left(\left|\frac{1+\sqrt{1-x}}{2}\right|+\left|\frac{1-\sqrt{1-x}}{2}\right|\right) \ ,
\end{equation}
where $z$ is a complex coordinate in the $2D$ plane, $z_{ij} \equiv z_i -z_j$ and $x=\frac{z_{12}z_{34}}{z_{13}z_{24}}$.
\subsection{Thermally perturbed two point functions}
As a first example, we consider a case in which we move away from the critical point by perturbing the Ising model with the energy operator, namely:
\begin{equation}
S=S_{\text{Ising}}+\lambda \int \mathcal{E}(z) d^2z \ . 
\end{equation}
Let us analyze the two point function $\langle \mathcal{E}(z_1) \sigma(z_2) \rangle_{\lambda}$. Assuming the the operator product expansion is valid outside the critical point we obtain:
\begin{equation}
\langle \mathcal{E}(z_1) \sigma(z_2) \rangle_{\lambda}=C^{\mathbbm{1}}_{\mathcal{E}\sigma}(z_{12}; \lambda )+C^{\mathcal{E}}_{\mathcal{E}\sigma}(z_{12}; \lambda) \langle\mathcal{E}\rangle_{\lambda}+C^{\sigma}_{\mathcal{E}\sigma}(z_{12}; \lambda)\langle \sigma\rangle_{\lambda} \ .
\end{equation}
The previous correlator vanishes identically at $\lambda=0$ since it contains just one insertion of $\sigma$. Moreover, by using the relation \eqref{confina}, one can easily prove that $C^{\mathbbm{1}}_{\mathcal{E}\sigma}(z_{12}; \lambda)$ and 
$C^{\mathcal{E}}_{\mathcal{E}\sigma}(z_{12}; \lambda)$ vanishes identically at all order in perturbation theory, since the correlation functions that one has to compute always contain an odd number of insertion of $\sigma$. Eventually, we find that, outside the critical point, the correlator is proportional to expectation value of $\sigma$, namely
\begin{equation}\label{propsigma}
\langle \mathcal{E}(z_1) \sigma(z_2) \rangle_{\lambda}=C^{\sigma}_{\mathcal{E}\sigma}(z_{12}; \lambda)\langle \sigma\rangle_{\lambda} \ ,
\end{equation}
and is non-zero only in the magnetically ordered phase. Expanding \eqref{propsigma} up to the first order in the coupling $\lambda$ we find:
\begin{equation}\label{2ptexp}
\langle \mathcal{E}(z_1) \sigma(z_2) \rangle_{\lambda}=\left(C^{\sigma}_{\mathcal{E}\sigma}(z_{12}) +\lambda \partial_{\lambda}C^{\sigma}_{\mathcal{E}\sigma}(z_{12})+...\right)\langle \sigma\rangle_{\lambda} \ ,
\end{equation}
where the dots stand for higher order corrections. Regarding the expectation value of $\sigma$, by dimensional analysis we obtain:
\begin{equation}
\langle \sigma\rangle_{\lambda}=A_{\sigma} \lambda^{\frac{1}{8}} \ ,
\end{equation}
where $A_{\sigma}$ is a non-universal constant. To compute $ C^{\sigma}_{\mathcal{E}\sigma}(z_{12})$ we rely on the operator product expansion \eqref{OPE} and on the orthogonality of the 2-pt correlation functions at the critical point, namely:
\begin{equation}
C^{\sigma}_{\mathcal{E}\sigma}(z_{12})=\lim_{|z_3|\rightarrow \infty} \frac{\langle \sigma(z_1) \sigma(z_3) \mathcal{E}(z_2) \rangle}{\langle \sigma(z_1) \sigma(z_3) \rangle}=\lim_{|z_3| \rightarrow \infty}\frac{|z_{13}|}{2|z_{12}||z_{23}|}=\frac{1}{2 |z_{12}|} \ .
\end{equation}
In order to evaluate the second term of the parenthesis in \eqref{2ptexp} we use the relation \eqref{con1} derived in the previous Section. Specifically, by setting $X=\sigma$ we obtain:
\begin{multline}\label{secondtherm}
\partial_{\lambda} C^{\sigma}_{ \mathcal{E}\sigma}(z_{12} ) \lim_{|z_2|\rightarrow \infty} \langle\sigma(z_2) \sigma(z_4) \rangle=
\lim_{|z_4|\rightarrow \infty} \int_{|z_3|<|z_4|} d^2 z_3 \left[\langle  \sigma(z_2) \sigma( z_4) \mathcal{E}(z_1) \mathcal{E}(z_3)\rangle\right.\\
\left. -C^{\sigma}_{\sigma \mathcal{E}}(z_{12})\langle  \sigma(z_2) \sigma(z_4)\mathcal{E}(z_3) \rangle -C^{\sigma_1}_{\sigma \mathcal{E}}(z_{12})\langle  \sigma_1(z_2) \sigma(z_4) \mathcal{E}(z_3) \rangle \right]
\end{multline}
where $\sigma_1 \equiv L_{-1} \bar{L}_{-1} \sigma$ and $L_n$ are the Virasoro operators.

The last term in \eqref{secondtherm} does not contribute to the integral. Regarding the other two terms, after dividing for $\langle\sigma(z_2) \sigma(z_4) \rangle$, performing the $|z_4|\rightarrow \infty$ limit and setting $z_1=0$ and $z_2 = z$ \footnote{This can be done due to conformal invariance at the critical point.} we obtain:
\begin{equation}
\partial_{\lambda} C^{\sigma}_{ \mathcal{E}\sigma}(z)=\int d^2 z_3 \frac{z_3 \bar{z}+\bar{z}_3 z}{2 |z_3||z||z-z_3|^2} \ .
\end{equation}
The integral has an UV divergence when $z_3$ approaches $z$. We regularise it by setting a cutoff at $|z_3|=|z|+\epsilon$. The final result is:
\begin{equation}
\partial_{\lambda} C^{\sigma}_{ \mathcal{E}\sigma}(z)=2 \pi \left(\log{\frac{2 |z|}{\epsilon}}-1\right) \ .
\end{equation}
To get rid of the UV regulator one has to consider the connected part of the correlator by subtracting to \eqref{2ptexp} the quantity $\langle \mathcal{E}(z_1)\rangle_{\lambda} \langle \sigma(z_2) \rangle_{\lambda}$. In order to do this, we need to compute the expectation value $\langle \mathcal{E}(z_1)\rangle_{\lambda}$. This has been done in \cite{Guida:1995kc} by noting that the action principle has to be valid, namely:
\begin{equation}\label{vevepsilon}
\partial_{\lambda}\langle \mathcal{E}(0)\rangle_{\lambda}= \int d^2 z_2 \langle \mathcal{E}(z_2)\mathcal{E}(0)\rangle_{\lambda}=-2 \pi \lambda \log{2 \pi \lambda e^{\gamma_E} \epsilon} \ ,
\end{equation}
where we refer to \cite{Guida:1995kc} for the details of the computation. Putting all together, we find:
\begin{equation}\label{connectedrr}
\langle \mathcal{E}(z_1) \sigma(z_2) \rangle_{\lambda}-\langle \mathcal{E}(z_1)\rangle_{\lambda} \langle \sigma(z_2) \rangle_{\lambda}=A_{\sigma} \lambda^{\frac{1}{8}} \left(\frac{1}{2 |z_{12}|}+ 2 \pi \lambda \left(\log{4 \pi \lambda |z_{12}|}-1+\gamma_E\right)+...\right) \ .
\end{equation}
Note that the computation \eqref{vevepsilon} is non perturbative, since, as explained in \cite{Guida:1995kc}, relies on the exact $\langle \mathcal{E}\mathcal{E}\rangle_{\lambda}$, which is known analytically outside the critical point for the special case of the 2D Ising model. As a consequence it is important to verify, as we have done, that the connected Green's function \eqref{connectedrr} does not depend on the UV regulator $\epsilon$. The expansion \eqref{connectedrr} is in perfect agreement with the exact result of the correlator found in  \cite{PhysRev.158.557}:
\begin{equation}
\langle \mathcal{E}(z_1) \sigma(z_2) \rangle_{\lambda}-\langle \mathcal{E}(z_1)\rangle_{\lambda} \langle \sigma(z_2) \rangle_{\lambda}= 2 \pi \lambda A_{\sigma} \lambda^{\frac{1}{8}} \int_{4 \pi \lambda |z_{12}|}^{\infty} ds \ s^{-2} e^{-s} \ .
\end{equation}
\subsection{Trapped Ising model}
\label{trapped}
As a second example, inspired by the analysis of \cite{Campostrini:2009ema}, we consider the 2D Ising model with a trap perturbation, namely:
\begin{equation}
S=S_{CFT}+\int d^2z \  U(z) \sigma(z) \ , \text{ with } U(z)=v^p |z|^p \equiv \rho |z|^p \ .
\end{equation}
In the previous expression $p$ ($\ge 2$) is a generic exponent of the trap potential while $\rho$ is the characteristic trap parameter. In what follows we will consider the effect of a small $\rho$ perturbation, namely the limit of large trap, on the two point functions of the Ising model. As explained in \cite{Campostrini:2009ema}, this might be of relevance in experimental studies of trapped critical systems such as cold atoms and Bose-Einstein condensates.

As noted in \cite{Campostrini:2009ema,Costagliola:2015ier}, by using renormalization group arguments, one can deduce the scaling behavior of the expectation value of the spin and the energy operator in the center of the trap:
\begin{equation}
\langle \sigma(0) \rangle_{\rho}=B_{\sigma} \rho^{\frac{\theta}{8}} \ , \qquad \langle \mathcal{E}(0) \rangle_{\rho}=B_{\mathcal{E}} \rho^{\theta} \ ,
\end{equation}
where the exponent $\theta=\frac{8}{15+8p}$ is the characteristic trap exponent, while $B_{\sigma}$ and $B_{\mathcal{E}}$ are non universal constants.

The perturbation that we are considering breaks translational invariance but preserves rotational symmetry. Hence, the fusion rule of the perturbed model remains the same as the unperturbed one and, if we fix one operator in the center of the trap, it is possible to rely on OPE and on the perturbative techniques described in the previous sections to compute the two point functions outside the critical point. The previous argument yields:
\begin{eqnarray}
\label{corrt1}
\langle \sigma(z_1) \sigma(0) \rangle_{\rho}&=& C^{\mathbbm{1}}_{\sigma \sigma} (z_1)+C^{\mathbbm{\mathcal{E}}}_{\sigma \sigma} (z_1) B_{\mathcal{E}} \rho^{\theta}+  \partial_{\rho} C^{\sigma}_{\sigma \sigma}(z_1) B_{\sigma} \rho^{\frac{\theta}{8}+1}+... \ ,\label{trap1}\\
\langle \mathcal{E}(z_1) \mathcal{E}(0) \rangle_{\rho}&=& C^{\mathbbm{1}}_{\mathcal{E}\mathcal{E}}(z_1)+   \partial_{\rho} C^{\sigma}_{\mathcal{E}\mathcal{E}} (z_1) B_{\sigma} \rho^{\frac{\theta}{8}+1}+... \ ,\label{trap2}\\
\label{corrt3}
\langle \sigma(z_1) \mathcal{E}(0) \rangle_{\rho}&=& C^{\sigma}_{\sigma \mathcal{E}} (z_1)B_{\sigma} \rho^{\frac{\theta}{8}}+\rho \ \partial_{\rho} C^{\mathbbm{1}}_{\sigma \mathcal{E}} (z_1)+  \partial_{\rho} C^{\mathcal{E}}_{\sigma \mathcal{E}} (z_1)B_{\mathcal{E}} \rho^{\theta+1}... \label{trap3} \ .
\end{eqnarray}
The last terms in \eqref{trap1}, \eqref{trap2} and \eqref{trap3} can be computed by using the conformal perturbation theory techniques described in the previous sections. Specifically, one has to evaluate the following integrals:
\begin{multline}
\label{inttrap1}
-\partial_{\rho} C^{\sigma}_{\sigma \sigma} (z_1) \lim_{|z_3| \rightarrow \infty} \langle \sigma(z_3) \sigma(0) \rangle=\lim_{|z_3| \rightarrow \infty} \int_{|z_2|<|z_3|} d^2z_2 \ |z_2|^p \Big[\langle \sigma(z_1) \sigma(z_2) \sigma(z_3) \sigma(0) \rangle\\
-C^{\mathbbm{1}}_{\sigma \sigma}(z_{1}) \langle \sigma(z_2) \sigma(z_3) \rangle-C^{\mathcal{E}}_{\sigma \sigma}(z_{1}) \langle \sigma(z_2) \sigma(z_3) \mathcal{E}(0) \rangle \Big] \ ,
\end{multline}
\begin{multline}
\label{inttrap2}
-\partial_{\rho} C^{\sigma}_{\mathcal{E}\mathcal{E}}(z_1) \lim_{|z_3| \rightarrow \infty} \langle \sigma(z_3) \sigma(0) \rangle=\\
\lim_{|z_3| \rightarrow \infty} \int_{|z_2|<|z_3|} d^2z_2 \ |z_2|^p \Big[\langle \sigma(z_2) \sigma(z_3)\mathcal{E}(z_1) \mathcal{E}(0) \rangle-C^{\mathbbm{1}}_{\mathcal{E} \mathcal{E}}(z_1) \langle \sigma(z_2) \sigma(z_3) \rangle \Big] \ ,
\end{multline}
\begin{multline}
\label{inttrap3}
- \partial_{\rho} C^{\mathbbm{1}}_{\sigma \mathcal{E}} (z_1)=
 \int d^2z_2 \ |z_2|^p \Big[\langle \sigma(z_1) \sigma(z_2) \mathcal{E}(0) \rangle-C^{\sigma}_{\sigma \mathcal{E}}(z_1) \langle\sigma(z_2) \sigma(0) \rangle \Big] \ ,
\end{multline}
\begin{multline}
\label{inttrap4}
- \partial_{\rho} C^{\mathcal{E}}_{\sigma\mathcal{E}}(z_1) \lim_{|z_3| \rightarrow \infty} \langle \mathcal{E}(z_3) \mathcal{E}(0) \rangle=\\
\lim_{|z_3| \rightarrow \infty} \int_{|z_2|<|z_3|} d^2z_2 \ |z_2|^p \Big[\langle \sigma(z_1) \sigma(z_2)\mathcal{E}(z_3) \mathcal{E}(0) \rangle-C^{\sigma}_{\sigma \mathcal{E}}(z_1) \langle \sigma(z_2) \mathcal{E}(z_3)\sigma(0) \rangle\\-C^{\sigma^1}_{\sigma \mathcal{E}}(z_1) \langle \sigma(z_2) \mathcal{E}(z_3)\sigma^1(0) \rangle \Big]
\end{multline}
A detailed discussion on how to treat the last expressions can be found in Appendix \ref{integrals}. Here we outline the results in the case of harmonic trap ($p=2$), which is the relevant one in typical experimental setups \cite{Campostrini:2009ema}:
\begin{eqnarray}
\partial_{\rho}C^{\sigma}_{\mathcal{E}\mathcal{E}}(z_1,p=2)&=&\frac{|z_1|^2}{64} \pi  (3+2 \gamma_E -2 \log (4))\\
\partial_{\rho} C^{\mathbbm{1}}_{\sigma \mathcal{E}}(z_1,p=2)&=&-|z_1|^{\frac{11}{4}} \ \frac{\pi  \cot \left(\frac{\pi }{8}\right) \Gamma
   \left(\frac{11}{8}\right)^2}{8 \Gamma
   \left(\frac{23}{8}\right)^2}\\
\partial_{\rho} C^{\mathcal{E}}_{\sigma \mathcal{E}}(z_1,p=2)&=&-|z_1|^{\frac{15}{4}} \ \frac{961 \Gamma \left(-\frac{19}{8}\right)^2 \Gamma
   \left(\frac{7}{8}\right)^2}{2048 \sqrt{2} \pi }\\
\partial_{\rho} C^{\sigma}_{\sigma \sigma}(z_1,p=2)&=&-|z_1|^{\frac{15}{4}} 0.00153398
\end{eqnarray}

In order to make contact with the experiments it is useful to express the final result in Fourier transform. In order to do this one has to keep in mind that for $n \neq  2 \mathbb{Z}$ (see \emph{e.g.} \cite{gelfand}):
\begin{equation}\label{222}
\mathcal{F}[|z|^n](|q|)=\frac{2^n \ n \ \Gamma \left(\frac{n}{2}\right)|q|^{-n-2}}{\Gamma 	\left(-\frac{n}{2}\right)} \ .
\end{equation}
while
\begin{equation}\label{333}
\mathcal{F}[|z|^2](q,\bar{q})=-2 \pi \partial_q \partial_{\bar q} \delta(q) \delta(\bar q) \ , \qquad \mathcal{F}[|z|^{-2}](|q|)= -\log \frac{|q|}{2 \mu}-\gamma_E\ ,
\end{equation}
where $\mu$ is a renormalization parameter. The last equivalence in \eqref{333} has been obtained by expanding \eqref{222} around the singular point $n=-2$ and taking the finite part.

Eventually, the Fourier transform of the correlators \eqref{corrt1}-\eqref{corrt3} is:
\begin{eqnarray}
\langle \sigma(q) \sigma(-q) \rangle_{\rho}&=&\frac{1}{|q|^{23/4}}\left(-0.3908 B_{\mathcal{E}}  \rho ^{\theta }|q|^{3}-0.01607 B_{\sigma} \rho ^{ \theta/8 +1}+0.2432|q|^4\right)\\
\langle \mathcal{E}(q) \mathcal{E}(-q) \rangle_{\rho}&=&-\log \frac{|q|}{2 \mu}-\gamma_E-0.42620 B_{\sigma} \rho^{\frac{\theta}{8}+1}  \partial_q \partial_{\bar q} \delta(q) \delta(\bar q)\\
\langle \sigma(q)  \mathcal{E}(-q) \rangle_{\rho}&=&-\frac{1.8026 }{q^{23/4}}B_{\mathcal{E}} \rho ^{\theta+1}+\frac{B_{\sigma} \rho ^{\theta/8 }}{q}-\frac{1.38552  }{q^{19/4}}\rho
\end{eqnarray}

\subsection{Magnetically perturbed three-point functions}
As we have pointed out in the first Section, CPT can be also applied to $3$-point correlators\footnote{See \cite{Caselle2006b} for an example using the Potts model.}. In what follows we will analyze the 2D Ising model perturbed by a magnetic field $h$, namely
\begin{equation}
S=S_{\text{Ising}}+h\int \sigma(z) d^2z \  .
\end{equation}

We will apply the conformal perturbation theory techniques described in Section \ref{CPT} to the study of 3-point correlation functions outside the critical point. Specifically we will analyze $\langle\sigma (z_1 ) \sigma (z_2 ) \mathcal{E} (z_3 )\rangle_h$. The assumption that the Operator Product Expansion is valid outside the critical point implies:
\begin{equation}
\langle\sigma (z_1 ) \sigma (z_2 ) \mathcal{E} (z_3 ) \rangle _h =  C^{1}_{\sigma\sigma\mathcal{E}}(z_1,z_2,z_3;h)+ C^{\sigma}_{\sigma\sigma\mathcal{E}}(z_1,z_2,z_3;h) \langle \sigma \rangle_h  +
C^{\mathcal{E}}_{\sigma\sigma\mathcal{E}}(z_1,z_2,z_3;h) \langle\mathcal{E} \rangle_h \ ,
\end{equation}
The leading correction to the correlation function comes from the term proportional to $\langle\mathcal{E} \rangle_h$, namely:
\begin{equation}
\langle\sigma (z_1 ) \sigma (z_2 ) \mathcal{E} (z_3 ) \rangle _h =  C^{\mathbbm{1}}_{\sigma\sigma\mathcal{E}} (z_1,z_2,z_3) + C^{\mathcal{E}}_{\sigma\sigma\mathcal{E}}(z_1,z_2,z_3)  \langle\mathcal{E} \rangle_h + .....
\end{equation}
In order to evaluate the previous expression we need the quantities $ C^{\mathbbm{1}}_{\sigma\sigma\mathcal{E}} (z_1,z_2,z_3) $ and $C^{\mathcal{E}}_{\sigma\sigma\mathcal{E}}(z_1,z_2,z_3) $, as well as the expectation value $\langle\mathcal{E} \rangle_h$. Dimensional analysis tell us that the latter one is given by:
\begin{equation}
\langle\mathcal{E} \rangle_h= A_{\mathcal{E}}h^{\frac{8}{15}} \ ,
\end{equation}
where $A_{\mathcal{E}}$ is a non-universal constant.
Moreover, by definition, $C^{\mathbbm{1}}_{\sigma\sigma\mathcal{E}} (z_1,z_2,z_3) $ is given by:
\begin{equation}
C^{\mathbbm{1}}_{\sigma\sigma\mathcal{E}} (z_1,z_2,z_3) =\langle\sigma(z_1) \sigma(z_2) \mathcal{E}(z_3) \rangle=\frac{|z_{12}|^{\frac{3}{4}}}{2|z_{13}||z_{23}|} \ .
\end{equation}
where we have used \eqref{4pising}. Finally, in order to compute  $C^{\mathcal{E}}_{\sigma\sigma\mathcal{E}}(z_1,z_2,z_3) $ we rely, as in the previous subsection, on operator product expansion \eqref{OPE} together with the orthogonality of the 2-pt functions at the critical point. Eventually, in the limit $|z_4| \to  \infty$,  and using the equations \eqref{wilsoneqsing} and \eqref{4pising}, we obtain:
\begin{equation}
 C^{\mathcal{E}}_{\sigma\sigma\mathcal{E}}(z_1,z_2,z_3) =\lim_{|z_4| \rightarrow \infty}\frac{\langle\sigma (z_1) \sigma (z_2) \mathcal{E}(z_3) \mathcal{E}(z_4)\rangle}{\langle\mathcal{E} (z_3) \mathcal{E}(z_4)\rangle} =\lim_{|z_4| \rightarrow \infty} \frac{|z_{12} ( z_{32}+z_{42})-2 z_{32} z_{42}|^2}{4 |z_{42} z_{32} z_{41}z_{31}| |z_{12}|^{\frac{1}{4}}} \ ,
\end{equation}
Performing the $|z_4| \to  \infty$ limit we get:
\begin{equation}
C^{\mathcal{E}}_{\sigma\sigma\mathcal{E}}(z_1,z_2,z_3) = \frac{1}{4} \frac{|z_{12}-2 z_{32}|^2}{|z_{32} z_{31}| |z_{12}|^{\frac{1}{4}}}
. \end{equation}
Putting all together, we finally obtain:
\begin{equation}
\langle\sigma (z_1 ) \sigma (z_2 ) \mathcal{E} (z_3 ) \rangle _h  = \frac{1}{2}\frac{|z_{12}|^{\frac{3}{4}}}{|z_{13} z_{23}| } \left ( 1 + A_{\mathcal{E}} h^{8/15} \frac{|z_{13}+z_{23}|^2}{2|z_{12}|} +...\right ) \ ,
\end{equation}
where the dots stands for higher order correction in the magnetic field $h$. Finally we note that in the limit $|z_3 | \gg |z_1 |\ , \ |z_2 |\gg|z_1|$ the following operator product expansion relation holds:
\begin{equation}
\langle\sigma (z_1 ) \sigma (z_2 ) \mathcal{E} (z_3 ) \rangle_h = \langle\sigma (z_1 ) \sigma (z_2 )\rangle_h  \langle\mathcal{E}(z_3) \rangle_h = \frac{ A_{\mathcal{E}} h^{\frac{8}{15}}}{|z_{12}|^{\frac{1}{4}}}\ .
\end{equation}

\section{Conclusions}
In this note we have shown that in order to get infrared finite correlation functions near a critical point
one has to modify the traditional approach to perturbation theory. We have illustrated the correct approach by giving various non trivial examples of its application. Specifically, we have applied the method to compute the corrections due to several kind of relevant perturbations of the 2D Ising model not yet considered in the literature. In the special 2D case the computation of the VEVs can be performed relying only on the knowledge of the CFT model using the truncated conformal space (TCS) approach \cite{Guida:1997fs}, and this simplify the discussion. However it is worth mentioning that the method illustrated in Section 2 is valid in any space-time dimension and it has already been applied to the $3D$ Ising model \cite{Caselle:2015csa, Caselle:2016mww,Komargodski:2016auf}, even though, at present, in the 3D case the computations of VEVs relies only on numerical simulations.

As a first example, we have considered the perturbation due to the energy operator. In particular we have proven that, in this case, the conformal perturbation theory result for the correlator $\langle \sigma \mathcal{E} \rangle$ totally agrees with the exact correlator in the broken phase obtained in \cite{PhysRev.158.557}. This might be seen as a first step in the direction of applying the method to higher dimensional models which exhibit a spontaneous symmetry breaking, as the $3D$ $O(N)$ model, where recent conformal bootstrap computations of the Wilson coefficient at the critical point \cite{Kos:2016ysd} may render the method described here applicable. In this case, the dynamics of Goldstone bosons might be relevant, and recently it has been pointed out in \cite{Lucas:2016fju} that they might be the responsible for CPT to not reproduce correctly logarithmic terms in the current-current correlator.   The method that we have illustrated just relies on the validity of the OPE outside the critical point. Since, as observed in \cite{Novikov:1984rf}, the OPE should also be valid in case of spontaneous symmetry breaking, it is worth to analyze if it is possible, by applying this method, to reproduce the correct result for the current-current correlator in the $O(N)$ model. \footnote{Another interesting application of the present method is the analysis of the transport properties near the critical point in the presence of spontaneous symmetry breaking of translations. In this case the results can be compared with the analogous holographic computation (see \emph{e.g.} \cite{Amoretti:2017xto,Amoretti:2016bxs,Amoretti:2016cad,Amoretti:2017tbk,Amoretti:2012kb}).}

As a second example, we have considered the perturbation due to the presence of a trap. This example might be of relevance both from the experimental and theoretical point of view. From the experimental side, the present analysis might be useful for comparing the results with experimental data on trapped systems such as cold atoms \cite{trap1,trap2,trap3,trap4}. From the theoretical point of view, it is worth to mention that the trap case has some similarities with considering the system at finite temperature. In fact, the temperature acts as a box in the euclidean time direction, so that the VEV will be modified, but the Wilson coefficients will remain unchanged. In this direction the method could be applied to compute the response functions near a quantum critical point at zero  or finite temperature \cite{Lucas:2016fju}.

With the third example, we have proven that the method can be extremely powerful also in computing correction to the nth-point correlation functions. In particular, we have analyzed the corrections to the three point function $\langle \sigma \sigma \mathcal{E}\rangle$ due to the presence of an external magnetic field. As a possible future direction, the present analysis combined with recent conformal bootstrap results, might be useful in order to get more insight in higher dimensional CFTs outside the critical point.

The present method can be also applied in cases where the critical model is perturbed with more than one relevant operator\footnote{See \emph{e.g.} \cite{Behan:2017dwr,Behan:2017emf} for other methods to treat CFTs perturbed by relevant operators.}. As an example, it would be interesting to apply it to the Ising model perturbed with both the magnetic field and reduced temperature in order to analyze the interplay of scales due to the coexistence of two kind of perturbations. Finally, results from CPT can be useful in connection with sum rules  as in the QCD case \cite{Shifman:2010zzb}, or in the quantum critical case \cite{Lucas:2016fju, Lucas:2017dqa}.
\begin{acknowledgements}
A special thanks goes to Michele Caselle, Slava Rychkov and William Witczak-Krempa for providing comments on a  preliminary version of the present paper. We would like also to thank Daniele Musso for useful conversations.
Nicodemo Magnoli thanks the support of INFN Scientific Initiative SFT: Statistical Field Theory, Low-Dimensional Systems, Integrable Models and Applications.
\end{acknowledgements}
\appendix
\section{Mellin transform and asymptotic properties of integrals}
\label{integrals}
The structure of the integrals that we have to calculate is of the following form:
\begin{equation}\label{integral}
I(m) = \int d^2 z \Theta (m |z| ) g(z) \ ,
\end{equation}
where $\Theta( m |z| ) = e^{-m|z|}$ is an infrared cutoff function which can be eventually set to 0 at the end of the calculation. In order to obtain the asymptotic ($m \sim 0$) expansion of the integral, we will make use of the Mellin transform of $I(m)$. Assuming that the leading behaviors of $I(m)$ are $m^{a}$ when $m \rightarrow 0$ and $m^{-b}$ when $m \rightarrow \infty$, one can define the Mellin transform $\tilde I (s)$ on the strip $-a < Re(s) < b$ in the complex $s$ plane  as:
\begin{equation}
\tilde{I} ( s ) = \int_0^\infty \frac{dm}{m} m^s I( m ) \ . 
\end{equation}
It is well known, see (\cite{wong2001asymptotic}), that the poles of the Mellin transform are in one to one correspondence with the asymptotic expansion of the original function at $m=0$. In fact one can write:
\begin{equation}
I(m) = \sum _i \left. Res( \tilde{I} (s) ) m^{-s} \right|_{s = -a_i} \ ,
\end{equation}
where $a_1 \equiv a < a_2<...$ are the powers of $m$ in the asymptotic expansion of $I(m)$ at $m \sim 0$.
This equation tells us that, if the infrared counterterms do not give any finite contribution, then we can get the corrections to the Wilson coefficients by taking the residue of the perturbative expansions at $s = 0$.

The Mellin transform previously described is particularly useful for our purposes if one notes that, by making use of the convolution theorem, it is possible to express the integral \eqref{integral} as:
\begin{equation}
\tilde{I} (s) = \Gamma (s) \tilde{g} (1-s) \ ,
\end{equation}
where
\begin{equation}
\tilde {g} (1-s) = \int d^2 z |z|^{-s} g(z) \ ,
\end{equation}
is essentially the Mellin transform of $g$ up to angular coefficients. Eventually, if one knows how to compute the Mellin transform of $g$, the integrals necessary to evaluate the derivative corrections of the Wilson coefficients are easily done.
In our case, all the integrals have the following form:
\begin{equation}\label{inteform}
I(m ; x) = \int d^2 z e^{-m |z|} |z|^{2\alpha } |z-x|^{2 \gamma} |z-1|^{2 \beta}
\end{equation}
The Mellin transform of this integral has been computed in \cite{Vladimir}, and is given by:
\begin{equation}
\tilde{I} (s;x) = \Gamma (s) D(\alpha -s/2, \beta , \gamma ,x )\ ,
\end{equation}
where the integral
\begin{equation}
D( a,b,c,x ) =  \int d^2  |z|^{2 a } |z-x|^{2 c} |z-1|^{2 b}
\end{equation}
can be expressed as:
\begin{equation}
D(a,b,c,x) = \frac{S(a) S(c)}{S(a+c)} |I_{0x}|^2 +\frac{S(b) S(a+b+c)}{S(a+c)} |I_{1\infty}|^2 \ .
\end{equation}
In the last expression, $S(a) \equiv sin(\pi a)$ and
\begin{equation}
I_{0x} \equiv x^{1+a+c} \frac{\Gamma (a+1) \Gamma (c+1) }{\Gamma (a+c+2)}   {}_2 F_1 (-b,a+1,a+c+2,x) \ ,
\end{equation}
\begin{equation}
I_{1\infty} \equiv \frac{\Gamma (-a-b-c-1) \Gamma (b+1) }{\Gamma (-a-c)} {}_2 F_1 (-a-b-c-1,-c,-a-c,x) \ ,
\end{equation}
where ${}_2 F_1 $ is the Hypergeometric Gauss function.
\subsection{Integrals for the trapped Ising model}
In this subsection we sketch the form of the integrals relevant to compute the perturbations of the trapped Ising model analyzed in Section \eqref{trapped}. These integrals can be evaluated by using the Mellin transform technique previously described. In this respect, it is sufficient to consider just the first terms in the integrals \eqref{inttrap1}-\eqref{inttrap4}. Eventually,  considering a generic value of $p$, and after evaluating the $|z_3|\rightarrow0$ limit, one has to compute the following integrals:
\begin{eqnarray}
\partial_{\rho}C^{\sigma}_{\mathcal{E}\mathcal{E}}(z_1)&=&-|z_1|^p \int d^2 w |w|^{p-1} |w-\frac{1}{2}|^2|w-1|^{-1} \ ,\\
\partial_{\rho} C^{\mathbbm{1}}_{\sigma \mathcal{E}}(z_1)&=&-\frac{|z_1|^{p+\frac{3}{4}}}{2}\int d^2 w |w|^{p-1}|w-1|^{\frac{3}{4}} \ ,\\
\partial_{\rho} C^{\mathcal{E}}_{\sigma \mathcal{E}}(z_1)&=&-\frac{|z_1|^{p+\frac{7}{4}}}{4}\int d^2 w |w|^{p-1}|w-1|^{-\frac{1}{4}}|w+1|^2 \ ,\\
\partial_{\rho} C^{\sigma}_{\sigma \sigma}(z_1)&=&-2|z_1|^{p+\frac{7}{4}} \int d^2w |w|^{\frac{3}{2}}|w-1|^{-(p-s+4)}|w+1|^{-(p-s+3)} \ .
\end{eqnarray}
As one can easily see, the previous integrals are of the form \eqref{inteform}, and can be evaluated by using the Mellin technique transform described  previously.

\end{document}